\documentclass[aps,prl,amsmath,amssymb,twocolumn,10pt]{revtex4-1}
\usepackage[noconfigs,english]{babel}
\usepackage{amssymb}
\usepackage{amsmath}
\usepackage{bm}
\usepackage{graphicx}
\usepackage{amsfonts}
\usepackage{latexsym}
\usepackage{SIunits}
\usepackage{cancel}
\usepackage[normalem]{ulem} 

\usepackage{color}
\definecolor{mygreen}{rgb}{0, 0.6, 0}

\usepackage[extension=xxx]{hyperref}
\def\be{\begin{equation}}   \def\ee{\end{equation}}
\def\eq#1{{Eq.(\ref{#1})}}    \def\fig#1{{Fig.\ref{#1}}}

\begin{document}
\title{Model of lamellipodium initiation during cell spreading}
\date{\today}
\author{Mathieu Dedenon \& Pierre Sens}%
\affiliation{Institut Curie, PSL Research University, CNRS, Physical Chemistry Curie, F-75005, Paris, France}
\affiliation{Sorbonne Universit\'{e}, CNRS, UMR 168, F-75005, Paris, France}

\begin{abstract}
Cell spreading requires a major reorganisation of the actin cytoskeleton, from a cortical structure to a lamellipodium where filaments are mostly parallel to the substrate. We propose a model inspired by the physics of nematic liquid crystals and fluid membranes, in which the coupling between actin mechanics, filaments orientation, and the local curvature of the cell membrane naturally yields the collective reorientation of actin filaments at the highly curved edge of a spreading cell. 
Filament orientation increases the traction force exerted by the frictional flow of polymerising actin on the substrate, creating a positive feedback loop between edge curvature, filament alignment, and traction force that promotes cell spreading.
We establish the condition under which this feedback leads to a full wetting transition, which we interpret as the initiation of a lamellipodium, and we uncover the existence of bi-stability between partial and full spreading which could trigger spontaneous cell polarization and lead to migration.
\end{abstract}
\maketitle

The spreading of eukaryotic cells on adhesive substrate occurs in distinct stages \cite{sheetz2008a}. The initial step is analogous to the wetting of passive liquid droplets\cite{degennes1985,frisch2002}, and follows a ``universal'' dynamics, where spreading is driven by a constant driving force (adhesion energy density) and balanced by a dissipative force associated to shape changes of the spreading cell \cite{cuvelier2007}. 
After this ``non-specific'' initiation \cite{pierres2003}, many cells generate a thin protrusion called lamellipodium made of a dense and polarised actin network pushing against the membrane edge \cite{small2002}. Lamellipodium expansion can be isotropic or anisotropic \cite{sheetz2004}, and a spontaneous symmetry breaking can lead to the motility of cells \cite{theriot2007} and cell fragments  \cite{verkhovsky1999}

Despite a large interest for lamellipodium structure and dynamics \cite{blanchoin2014,carlier2010}, a key process for cell spreading and crawling, the mechanisms involved in its initiation are much less understood and (surprisingly) rather unexplored. This process  requires a dramatic remodelling of the cell edge, from a smooth rounded shape to a thin and flat protrusion, and a reorganisation of the actin cytoskeleton, from a thin, unordered cortical layer to a polarised flat sheet. These two phenomena are clearly coupled. The structure of the actin network is influenced by the shape of the membrane from which it grows \cite{noireaux2000}, but it also influences the actin mechanical stress on the membrane, which modifies the shape \cite{boukellal2004}. The goal of this paper is to describe this coupling and investigate how it regulates cell spreading on flat substrates.
\begin{figure}[b]
\includegraphics[width=8.5cm]{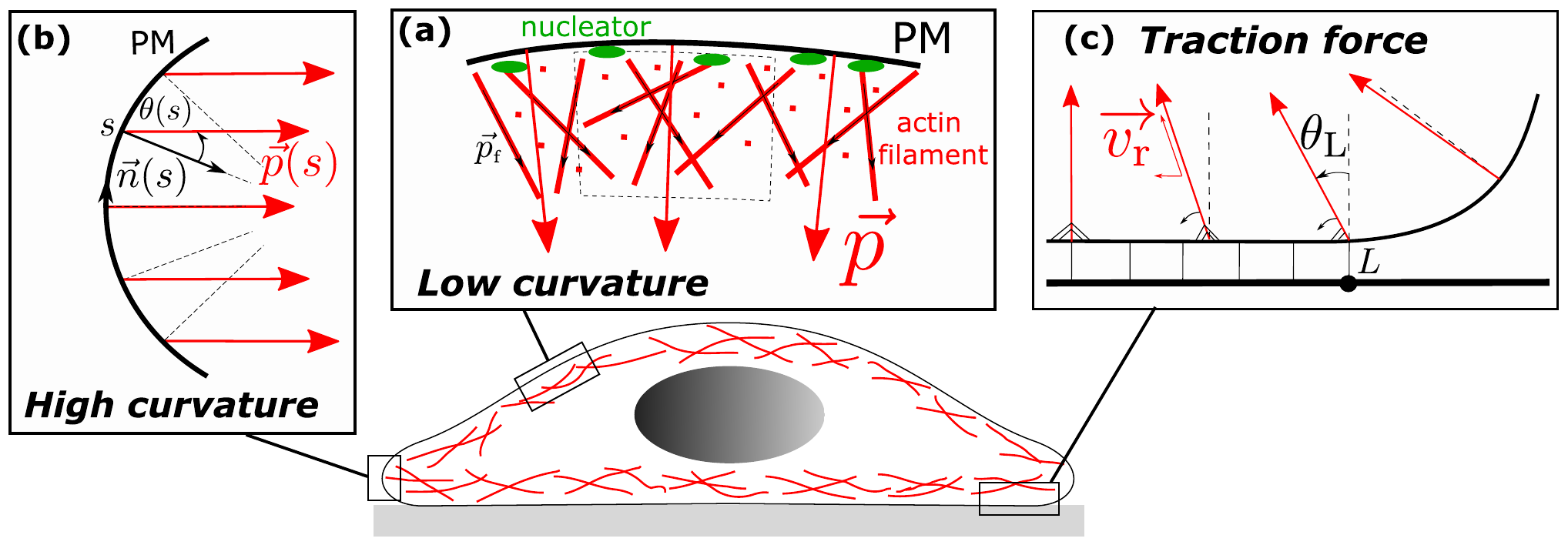}
\caption{Sketch of the model of cytoskeleton organisation in a spreading cell. 
\textbf{(a)}: The actin cortex is described by a polarization field $\vec{p}$ (the average on filaments polarization $\vec{p}_{\text{\tiny f}}$). Constant polymerization near plasma membrane and no preferred tangential direction orient $\vec{p}$ along the normal vector if curvature is negligible.
\textbf{(b)}: In regions of high curvature, the orientation of the polarisation field  $\vec{p}(s)$ deviates from the membrane's normal by angle $\theta(s)$ to release the mechanical stress associated with gradients of actin retrograde flow. \textbf{(c)} The interaction of the actin retrograde flow with adhesion molecules on the substrate creates a friction (traction) force, proportional to the horizontal projection of the flow.}
\label{fig1_sketch}
\end{figure}

The cell cortex is a thin ($\unit{100}{\nano\meter}-\unit{1}{\mu\meter}$) layer of  filamentous actin network, myosin motors, and actin-regulatory proteins underlying the plasma membrane (PM) of most eukaryotic cells lacking a cell wall \cite{salbreux2012,paluch2018}, whose main function is to regulate cell shape \cite{kapustina2013}. It has been described as an active viscoelastic material  within which acto-myosin contraction generates an active cortical tension able to drive morphological changes \cite{julicher2007}. Its viscous nature stems from the constant actin turnover over time scales of order $1-10\ {\rm min}$ due to polymerising factors at the PM (Arp$2/3$ and Formins), and depolymerising factors within the cortex \cite{fritzsche2013,paluch2014}.

The orientation of cortical actin filament in the plane of the membrane has been introduced previously, {\em e.g.} to account for the formation of the contractile ring during cytokinesis \cite{salbreux2009,turlier2014}. To describe the cortex-to-lamellipodium transition, one must account for the filament's orientation {\it normal to the membrane}, and in particular how this couples to membrane curvature. To this aim, we introduce a minimal model where the cortex structure is defined by a single order parameter; a polarization field $\vec{p}$ which is the mesoscopic average of filaments polarization $\vec{p}_{\text{\tiny f}}$, as shown in \fig{fig1_sketch}a.
On a flat membrane, there is no preferred direction tangential to the membrane and $\vec p$ is along the membrane's normal vector, even though actin filaments may be mostly  parallel to the PM \cite{morone2006,paluch2018}. Because of actin turnover, the polarisation field also represents the average {\it retrograde} flow of actin away from the PM. On a curved surface, visco-elastic stress with the growing actin layer tends to align neighbouring polarisation vectors  \cite{julicher2007}, see \fig{fig1_sketch}b. We write an effective elastic energy inspired by the Frank's energy density of nematic liquid crystals \cite{degennes1995}, where alignment of the actin flow translates to a term proportional to $|\partial_s\vec{p}(s)|^2$, whereas the natural tendency of the flow to align normal to the membrane translates into a generic anchoring term $(\vec{p}\cdot\vec{t})^2$. 

In what follows, we neglect the curvature of the cell surface in the azimuthal direction (along the cell edge) and study the spreading of a two-dimensional cell for simplicity. This approximation is valid in the limit where the radius of the adhered area is much larger than the radius of curvature of the cell edge. The polarisation vector $\vec{p}(s)$ may thus be replaced by its angle with normal vector $\theta(s)$ (see \fig{fig1_sketch}b). Including the membrane bending rigidity and surface tension from the classical  Helfrich-Canham energy \cite{helfrich1973}, we obtain the  full energy of the cell interface which, up to second order in membrane curvature $C$ and cortex orientation $\theta$, reads:
\be
\label{energy}
E_{\text{\tiny m}{\scriptscriptstyle\oplus}\text{\tiny c}}=\frac{1}{2}\int\mathrm{d}s\left\{2\gamma+\kappa C^2+k_\alpha\theta^2+\kappa_{\rm c}(C-\partial_s \theta)^2\right\}
\ee
Here, $s$ is the curvilinear coordinate along the membrane, $\gamma$ is the tension of the interface, which can include a contribution from cortical acto-myosin contraction, $\kappa$ is the membrane bending rigidity, $\kappa_{\rm c}$ is an effective bending rigidity for the cortex, which depends on the cortex's Young modulus, visco-elastic relaxation time, and polymerisation velocity \cite{julicher2007}, and $k_\alpha$ characterises the rigidity of the cortex' normal anchoring to the PM. Note that the effect of filament orientation on the cortical tension, which relies on myosin contraction of antiparallel actin filaments, is not taken into account in our simple model that includes a single orientation parameter $\theta$.

We first investigate the mechanical response of a weakly deformed interface, see the Supplementary Information (S.I.) for more detail. We use the Monge representation where $h(x)$ is the deformation with respect to the flat state, with $ds\simeq1+\partial_x h^2/2$ and $C(x)\simeq\partial_x^2 h(x)$ for small deformation: $\partial_x h\ll 1$. The Euler-Lagrange equations derived from the interface energy \eq{energy} are:
\begin{eqnarray}
&\lambda^2\left[(1+e)\partial_x^2C-e\partial_x^3\theta\right]=C\ ,\  \lambda_{\rm c}^2\left(\partial_x^2\theta-\partial_xC\right)=\theta\cr
&\text{with}\quad \lambda\equiv\sqrt{\kappa/\gamma},\quad \lambda_{\rm c}\equiv\sqrt{\kappa_{\rm c}/k_\alpha},\quad e\equiv\kappa_{\rm c}/\kappa
\label{el}
\end{eqnarray}
where $\lambda$ and $\lambda_{\rm c}$ are the characteristic length scales of membrane deformation and cortical orientation, and $e$ is the ratio of the cortical to membrane bending rigidities. In the quasi-static limit, when cortical orientation can adjust to membrane deformation, the interface elastic energy can be written in Fourier space as $E_{\text{\tiny m}{\scriptscriptstyle\oplus}\text{\tiny c}}=\int\mathrm{d}q\,\frac{K_q}{2}|h_q|^2$, with the scale-dependent stiffness
\be 
K_q=\gamma q^2\left\{1+(\lambda q)^2\left[1+\frac{e}{1+(\lambda_{\rm c} q)^2}\right]\right\}.
\label{fourier}
\ee
Comparing this response function to that of a pure membrane: $K_q^{\text{\tiny m}}=\gamma q^2\left\{1+(\lambda q)^2\right\}$, one sees that the membrane with cortex behaves as a pure membrane with a larger effective bending stiffness $\kappa_{\rm eff}=\kappa(1+e)$, with little reorganisation of actin polarisation ($\theta\simeq 0$), for small curvature ($\lambda_{\rm c} q\ll1$). For large curvature ($\lambda_{\rm c} q\gg1$), variations of the actin orientation closely follows the local membrane curvature ($\theta'\simeq C$), and the interface behaves as a pure membrane with a ``pseudo'' surface tension $\gamma_{\rm eff}=\gamma(1+\alpha)$, with $\alpha\equiv k_\alpha/\gamma$.

\begin{figure}[b]
\includegraphics[scale=0.4]{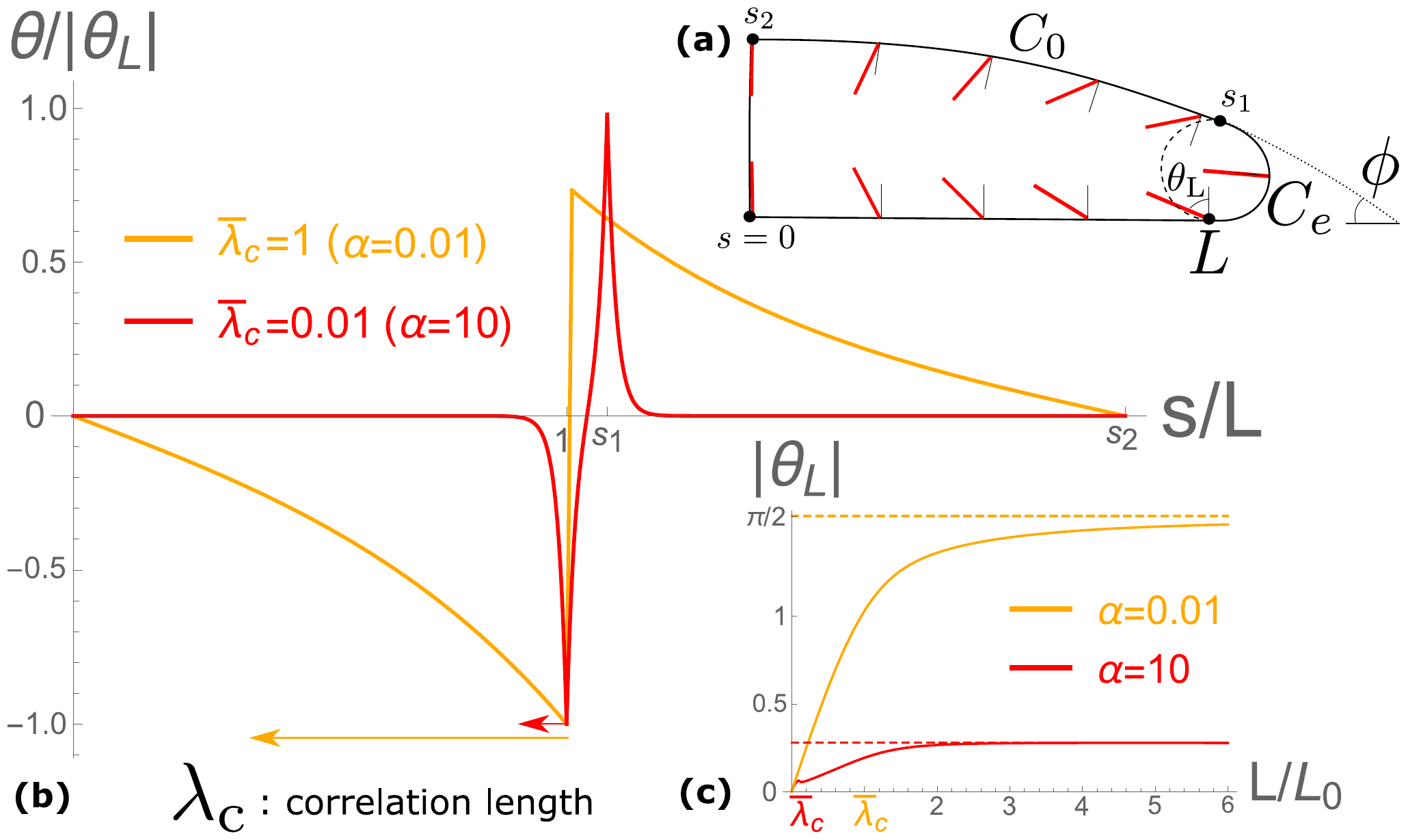}
\centering
\caption{{\bf (a):} Sketch of a cortical vesicle with parametrization of the membrane shape given by an adherent length $L$, a cap curvature $C_0$ and an edge curvature $C_{\rm e}$, with an illustrative cortical orientation profile. The effective wetting angle $\phi$ is indicated.
{\bf (b):} Normalized orientation field $\theta$ along the cell membrane as a function of arc-length $s$ for weak (orange) and strong (red) anchoring. The variables $C_0(L)$ and $C_{\rm e}(L)$ have been minimized for a fixed adherent length $L=1.8L_0$.
{\bf (c):} Cortical reorientation amplitude at contact point $|\theta_L|$ as a function of $L$. Parameters are  $\bar\lambda=10^{-2}$ and $e=100$.}
\label{fig2_orientation}
\end{figure}

Cell spreading bears strong similarities with the wetting of a liquid droplet \cite{frisch2002}. The static shape of a wetting droplet is related to the adhesion energy density $\epsilon$ and the interfacial tension $\gamma$ by the Young-Dupr\'e equation: $\gamma(1+\cos\phi)=\epsilon$ (where $\phi$ is the contact angle between the free surface and the substrate - see \fig{fig2_orientation})  \cite{degennes1985}.
Wetting does not occur ($\phi=\pi$) in the absence of adhesion ($\epsilon=0$), and a transition to full wetting ($\phi=0$) occurs when $\epsilon>2\gamma$.
For a spreading cell, the ``adhesion energy'' has a contribution from the traction force exerted by treadmilling actin filaments on the substrate by way of transient binding and unbinding of adaptor proteins such as talin or vinculin, linking cytoskeleton to integrins \cite{sheetz2003}.
In a well-developed lamellipodium, actin filaments orient parallel to the substrate ($\theta\sim\pi/2$). They exhibit an horizontal retrograde flow away from the cell edge due to their polarised polymerisation at the cell membrane, creating a traction force akin to a friction force \cite{sens2015}. However at the beginning of spreading the local retrograde flow is not parallel to the substrate, hence the force balance at the cell's leading edge is the integral over the adhered zone (of length $L$) of a friction force density proportional to the horizontal projection of the mean filament polarisation: $\sin{\theta(s)}\simeq \theta(s)$ (see \fig{fig1_sketch}c). At steady-state, this friction force is integrally transmitted to the cell edge, leading to the force balance:
\begin{equation}\label{force_balance}
F_{\text{\tiny m}{\scriptscriptstyle\oplus}\text{\tiny c}}=\partial_{\rm L}E_{\text{\tiny m}{\scriptscriptstyle\oplus}\text{\tiny c}}=\epsilon-\epsilon_{\rm f}\int_{\text{\tiny L}}\mathrm{d}s\,\sin\theta(s).
\end{equation}
which reduces to the Young-Dupr\'e relation in the absence of cytoskeleton and for negligible bending rigidity. Here, $\epsilon_{\rm f}\equiv\xi v_{\rm r}$ is the traction force parameter, that depends on the actin retrograde flow velocity $v_{\rm r}$ \cite{sheetz2006} and a friction parameter $\xi$ function of the density, stiffness, and binding kinetics of the adhesion proteins \cite{sabass2010,sens2013} (see S.I.).

Note that the interaction with the substrate can also modify the cortical parameters appearing in \eq{energy}. It is shown in the S.I. that the torque exerted by the adhesion molecules on the filaments promotes their parallel orientation by effectively reducing $k_\alpha$, but that this effect can be neglected for physiological parameters.

In the S.I.,  the force balance \eq{force_balance} is used to study the wetting by a membrane with cortex of a wedge of fixed angle. Wetting is limited by the increase of curvature as the membrane penetrates the wedge. For small angle, linearisation of the shape similar to the one leading to \eq{el} allows to obtain  implicit analytical equations for the penetration length. This illustrates the two main effects of the cortex: an increase of the bending rigidity of the interface hinders penetration at low membrane curvature, and an increase of the traction force due to filament reorientation promotes penetration at high curvature.

Below, we study the effect of the cytoskeleton on the spreading of a (2D) cell such as the one sketched in \fig{fig2_orientation}. The adhesion of cortex-free vesicles has been studied in depth \cite{seifert1990,lipowsky1991}. Compared to a wetting droplet, the definition of a contact angle is somewhat more complex for vesicles with a finite bending rigidity, since the edge curvature is finite and  equal to $\sqrt{\kappa/2\epsilon}$ \cite{seifert1990}. Nevertheless the Young-Dupr\'e relation still gives valuable information in the limit where the vesicle size is much larger than the membrane elastic length $\lambda$.

The thermodynamic ensemble for a fluid drop is that of constant interfacial tension, adhesion energy density and volume. Lipid vesicles  have a fixed membrane area and their membrane tension typically increases with spreading. As a consequence, the full wetting transition does not exist for vesicles, which exhibit a finit adhered area, or vesicle bursting if $\epsilon$ is larger than twice the membrane lysis tension \cite{lipowsky1991}. Contrarily to vesicles, cells regulate their mechanical properties through active processes. The tension of the cell membrane is regulated by membrane reservoirs \cite{sens2006,sinha2011}, and spreading cells also experience bursts of exocytosis that maintain the  membrane tension to an homeostatic value while increasing the cell membrane area \cite{sheetz2011,theriot2013}. Wether the cell volume changes during spreading is still a matter of debate, although recent findings suggest a small volume decrease over a long time scale \cite{guo2017,xie2018}.
As we are mostly concerned here with the initiation of the lamellipodium, we overlook the complexity of the biological regulations and choose the thermodynamic ensemble where interfacial tension, adhesion energy density and volume are constant. This allows for the existence of a full wetting state correlated with the reorientation of actin filaments, which we identify as the initiation of a lamellipodium. As we discuss below, this transition relies mostly on a local coupling between membrane curvature and cytoskeleton organisation, and should not depend much on the chosen ensemble.

The shape and cortical orientation of a spreading cell can be obtained for a given adhering length $L$ by minimizing the cell's mechanical energy \eq{energy} with respect to the local curvature $C(s)$ and cortical orientation $\theta(s)$. Here we use a simpler approach where the cell shape is parametrised by a flat region for $0\le s\le L$, an edge of constant curvature $C_{\rm e}$ for $L\le s\le s_1$, and a spherical cap of curvature $C_0$ for $s_1\le s\le s_2$, as sketched in \fig{fig2_orientation}a. This parametrisation compares well with the exact numerical solution for a simple vesicle without cortex (see S.I.). The local cortex orientation is given by $\lambda_c^2(\partial_s^2\theta-\partial_s C)=\theta$, as in \eq{el}, with appropriate boundary condition of torque balance - continuity of $(\partial_s\theta-C)$ - at curvature discontinuities ($s=L$ and $s=s_1$), and normal cortical orientation at the cell centre imposed by axial symmetry ($\theta(0)=\theta(s_2)=0$). The resulting energy is minimised with respect to $C_{\rm e}$ and $C_0$ for a given $L$.

In physiological conditions, one expects $\kappa\sim 10-25 k_{\rm B}T$ \cite{zhelev1994}, $\gamma\sim 10^{-5}-10^{-3}\unit{}{\newton/\meter}$ \cite{paluch2009}, such that $\lambda\lesssim\unit{100}{\nano\meter}$, and $\kappa_{\rm c}\sim 200-1000 k_{\rm B}T$ \cite{simson1998}. Choosing a characteristic cell size $L_0\sim\unit{10}{\micro\meter}$, we fix the dimensionless membrane length scale to the small value $\bar\lambda\equiv\lambda/L_0=10^{-2}$ and the dimensionless cortex rigidity to a large value $e\equiv \kappa_{\rm c}/\kappa=10^2$. We vary the ``anchoring strength'' $\alpha\equiv k_\alpha/\gamma$, which is more difficult to estimate from current data.

Representative cortical profiles are shown in \fig{fig2_orientation}b. Cortex reorientation is maximal near the cell edge ($s=L$) where curvature gradients are large. It decays over the length scale $\lambda_{\rm c}=\lambda\sqrt{e/\alpha}$ (\eq{el}). 
For large $L$, the maximal reorientation $\theta_L$ and the curvature of the edge $C_{\rm e}$ can be approximately obtained by requiring that the curvature saturates at the inverse length scale $C_{\rm e}\approx\sqrt{2\gamma/\kappa_{\rm eff}(C_{\rm e})}$, where $\kappa_{\rm eff}(C_{\rm e})$ is a scale dependent effective bending rigidity that decreases upon cortex reorientation, itself increasing with the curvature (as in \eq{fourier}, see S.I.). This leads to the asymptotes $\lambda C_{\rm e}\approx\sqrt{2/(1+e)}$ and  $\theta_L\approx-1/\sqrt{2\alpha}$ for strong anchoring $\alpha\gg1$, and $\lambda C_{\rm e}\approx\sqrt{2}$ and $\theta_L\approx-\pi/2$ for weak anchoring $\alpha\ll 1$, see \fig{fig2_orientation}c.

\begin{figure}[t]
\includegraphics[width=8.5cm]{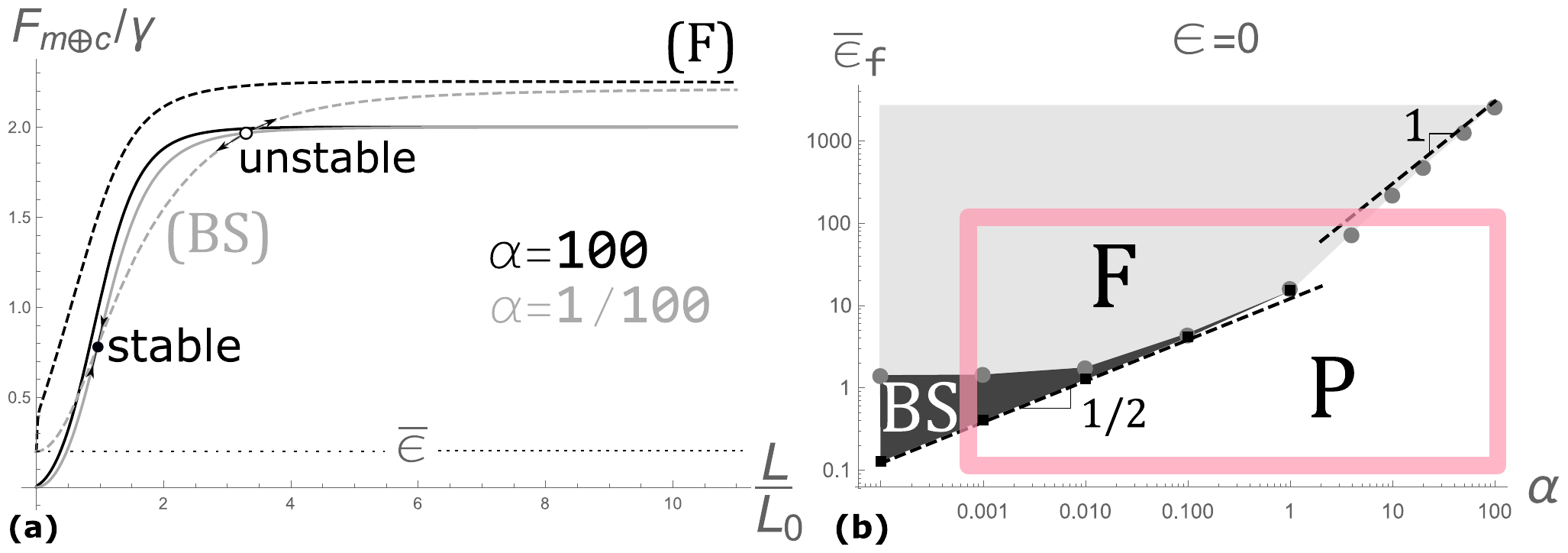}
\centering
\caption{{\bf (a)} Graphical resolution of the force balance equation \eq{force_balance}. The restoring force (solid lines) and the total traction force (dashed lines) are shown as a function of the adhered length  for weak (grey, with $\bar\epsilon_{\rm f}=1.3$) and strong (black, with $\bar\epsilon_{\rm f}=2750$) anchoring. Equilibrium states (P) correspond to these two forces being equal. Full spreading (F) is when the traction force exceeds the restoring force for $L\rightarrow\infty$. For weak anchoring, the system might show bi-stability (BS), with a (meta)stable and an unstable solution leading to full spreading. The passive adhesion force (dotted line) is $\bar\epsilon=0.2$. Other parameters are: $\bar\lambda=0.01$ and $e=100$.  {\bf(b)} Phase diagram in the $\alpha,\bar\epsilon_{\rm f}(=\epsilon_{\rm f}L_0/\gamma)$ phase space (log-log plot) for negligible non-specific adhesion ($\epsilon\ll\gamma$) showing the scaling behaviour predicted by \eq{fullspreading} and the extent of the bi-stability region.  The pink frame delimits the physiological accessible  range of parameters.}
\label{fig3_force}
\end{figure}

The force balance equation \eq{force_balance} is solved graphically on \fig{fig3_force}a by representing the membrane restoring force, and effective adhesion force as a function of the adhered length $L$. For a vesicle without cortex, the adhesion energy $\epsilon$ is constant and the restoring force increases (monotonically if $\bar\lambda\ll 1$, see S.I.) to the value $2\gamma$, with the full wetting condition $\epsilon>2\gamma$. With cortex, the restoring force is modified in the strong anchoring regime, due to an effectively more rigid membrane ($\lambda_{\rm eff}=\lambda\sqrt{1+e}$). It may possibly overshoot, but it saturates at the same value $2\gamma$. The traction force, on the other hand, increases to a value corresponding to the maximal reorientation of the cortex $\theta_L$ discussed above and shown in \fig{fig2_orientation}c.
The total effective adhesion force $\epsilon+f_{\rm f}$ includes the integrated traction force $f_{\rm f}$, which in the limit $L\gg \lambda_c$, reads $f^\infty_{\rm f}\sim\epsilon_{\rm f}\lambda_{\rm c}|\theta_L|$.
Hence, full wetting is possible when the friction parameter  $\epsilon_{\rm f}$ exceeds a threshold given by $\epsilon+f_{\rm f}^{\infty}\gtrsim 2\gamma$, with the following asymptotes in the strong and weak anchoring regimes:
\be
\epsilon_{\rm f}= \frac{2\gamma-\epsilon}{\lambda\sqrt{e}}g(\alpha)\quad,\quad  g(\alpha)\sim\begin{cases}
   \sqrt{\alpha}, & \text{if $\alpha\ll 1$}\\
   \alpha, & \text{if $\alpha\gg 1$}
  \end{cases}.
\label{fullspreading}
\ee
The full spreading transition is summarized on a phase diagram in \fig{fig3_force}b relating the threshold values of traction force amplitude $\epsilon_{\rm f}$ to the anchoring rigidity $\alpha$, in excellent agreement with the scaling law \eq{fullspreading}.

Another important characteristics of the traction force is the way it reaches the saturation value. For strong anchoring, the traction force density propagates a short distance $\lambda_c\ll L_0$ away front the cell edge, and the total traction force saturates  together with the edge curvature, when $L\sim L_0$.
For weak anchoring, and so long as $L<\lambda_c$, cortex reorientation propagates over the entire adhered region, and the net traction force scales linear with the adhered length: $f_{\rm f}\sim\frac{\pi}{4}\epsilon_{\rm f}L$ (for full cortex reorientation: $\theta_{L}\simeq\pi/2$). Crudely assuming that the membrane restoring force also scales linearly with the adhered length for small length: $F_{\text{\tiny m}{\scriptscriptstyle\oplus}\text{\tiny c}}\sim \gamma L/L_0$, the force balance could admit two stationary solutions, one stable and one unstable,  if the traction force increases slower than the restoring with the adhered length. This behaviour, illustrated on \fig{fig3_force}, should occur if  $\bar\epsilon_{\rm f}=\epsilon_{\rm f}L_0/\gamma< 1$. This simple argument, confirmed by numerical calculations, predicts that the phase diagram contains a region of bistability in the low anchoring regime, where a partial spreading state coexists with the full spreading state, see \fig{fig3_force}.

We now estimate the physiological range of the different parameters. The maximal traction force density  is of order $\epsilon_{\rm f}\simeq \rho_{\rm l}k_{\rm l}v_{\rm r}/k_{\rm off}$, where $\rho_{\rm l}$ is the density of ligand between actin and the substrate, $k_{\rm l}$ their stiffness, $k_{\rm off}$ their unbinding rate and $v_{\rm r}$ the actin retrograde flow velocity (see S.I. and \cite{sabass2010,sens2013}). With $\rho_{\rm l}\sim\unit{10}/{\micro\meter^2}$, $k_{\rm l}\sim\unit{1}{\pico\newton/\nano\meter}$, $k_{\rm off}\sim\unit{0.1-1}/{\second}$ and $v_{\rm r}\sim \unit{1-10}{\nano\meter/\second}$, we find $\epsilon_{\rm f}\sim\unit{10-10^3}{\pascal}$, which agrees well with the typical traction force density measured in spreading cell \cite{reinhartking2005,gardel2008}, and $\bar\epsilon_{\rm f}\sim 10^{-1}-10^2$. The anchoring parameter $\alpha=k_\alpha/\gamma$ is difficult to estimate, since it depends on the coupling between actin filaments and the membrane at the molecular scale, but also on the cortex structure (and hence is not independent of $\kappa_{\rm c}$). At the scaling level, the anchoring stiffness should satisfy $k_\alpha\simeq k_{\rm a} l_{\rm a}^2\rho_{\rm a}$, where $k_{\rm a}\sim\unit{1-10}{\pico\newton/\nano\meter}$ is a stiffness of membrane-actin anchors, $l_{\rm a}\sim\unit{3-10}{\nano\meter}$ is their characteristic length, and $\rho_{\rm a}\sim{10^2-10^3}/{\micro\meter}^2$ is the 2D density of anchoring points. This gives a broad range $k_\alpha\sim\unit{10^{-6}-10^{-3}}{\newton/\meter}$, or $\alpha\sim 10^{-3}-10^2$, which encompasses the three possible regimes: partial spreading, bistability, and full spreading - see \fig{fig3_force}.

In summary, we propose that there exists a mechanical feedback loop between the membrane curvature, cytoskeleton structure and traction force at the edge of spreading cells. High curvature drives actin filaments to orient parallel to the substrate, which increases the traction force, drives cell spreading and increases the edge curvature.  If the tension of the cell membrane remains fixed (regulated) during spreading, this can lead to full (unbounded) spreading. We argue that this corresponds to the initiation of a lamellipodium, the thin and flat protrusion of oriented actin filaments found at the edge of many spreading and crawling cells.

This transition is mostly controlled by two parameters (see \fig{fig3_force}): the ratio of traction force of a fully oriented lamellipodium over the membrane tension $\bar\epsilon_{\rm f}=\epsilon_{\rm f}L_0/\gamma$, and the ratio of the cortex anchoring strength over tension $\alpha=k_\alpha/\gamma$. The range of physiological values for these parameters corresponds to the region in the phase space where interesting things happen, namely the transition between partial and full spreading (lamellipodium initiation), with the possible coexistence of the two states when lamellipodium initiation is a nucleation event (\fig{fig3_force}b). The latter case may be indicative of anisotropic cell spreading since this suggests that lamellipodium initiation can be heterogeneously triggered by local fluctuations of the edge curvature. It also constitutes a possible mechanism for spontaneous symmetry breaking and crawling of spreading cells. Our model predicts that relatively weak variations of intracellular and extracellular parameters may modulate the cell spreading behaviour. For instance, we predict that decreasing the traction force parameter $\epsilon_{\rm f}$,  should trigger a transition from isotropic spreading (``full spreading'') to anisotropic spreading (``bistability'') (\fig{fig3_force}b). This transition has indeed been observed endothelial cells  upon reduction of the the substrate ligand density \cite{reinhartking2005}.

\begin{acknowledgements}
We thank Nir Gov and Jacques Prost for interesting discussions and insight.
\end{acknowledgements}

\bibliographystyle{unsrt}


\end{document}